\documentclass[twocolumn,showpacs,preprintnumbers,amsmath,amssymb]{revtex4}

\usepackage{graphicx}

\begin{document}

\title{Magnetoplasmons excitations in Graphene for filling factors $\nu \leq 6$}

  \author{Yu.A.\ Bychkov}
      \altaffiliation  {Also at L.D.Landau Institute for Theoretical Physics, Academy of Sciences
   of Russia, 117940 Moscow V-334, Russia }
  \author{G.\ Martinez}
  \affiliation{Grenoble High Magnetic Field Laboratory, CNRS, B.P.~166, 38042 Grenoble Cedex 9, France}


\date{\today}

\begin{abstract}
In the frame of the Hartree-Fock approximation, the dispersion of
magnetoplasmons in Graphene is derived for all types of
transitions for filling factors $\nu\leq 6$. The optical
conductivity components of the magnetoplasmon curves are
calculated. It is shown that the electron-electron interactions
lead to a strong re-normalization of the apparent Fermi velocity
of Graphene which is different for different types of transitions.

\end{abstract}

\maketitle



\section{ Introduction}

Graphene is a monolayer of Graphite with a band structure composed
of two cones located at two inequivalent corners \textbf{K} and
\textbf{K'} of the Brillouin zone at which conduction and valence
bands merge. This compound has recently received a large attention
because of the unusual sequence of quantum Hall states it reveals
\cite{Zhang,Novoselov}. In contrast to conventional
two-dimensional electron gas (C2DEG) which display a quadratic
dispersion law, Graphene exhibits a linear dispersion law
$E(\overrightarrow{p})= \pm v_{F}|\overrightarrow{p}|$ as a
function of the momentum $\overrightarrow{p}$ leading to a Dirac's
type Hamiltonian with a Fermi velocity $v_{F}$ replacing that of
the light. Different one-electron band structure models, not
including electron-electron interactions, lead to values
$v_{F}\simeq 0.86\times 10^{6} m/s$ with some variance, but this
is the value which will adopted in this report. This peculiar
dispersion law has  two important consequences in contrast to
C2DEG (see for instance \cite{Zheng}): (i) the wave functions have
a spinor type character and (ii) under a magnetic field \textbf{B}
applied perpendicular to the Graphene plane, the Dirac energy
spectrum evolves into Landau Levels (LL) with energies given by:
\begin{equation}
E_{n}= sgn(n)v_{F}\sqrt{2e\hbar B|n|}=sgn(n)E_{10}\sqrt{|n|}
\end{equation}
where $n$ scans all positive and negative integer values including
zero.

 Magnetoplasmons (MP), in a two-dimensional electron gas,
are excitations between LL,  known to be described in terms of
excitonic transitions due to electron-electron interactions (EEI):
they reveal a specific dispersion as a function of the two
dimensional wave vector $\vec{k}$ of the exciton. For a C2DEG, the
theory, derived in the frame of the Hartree-Fock (HF)
approximation, has been first developed \cite{KH,KM} for integer
values of the filling factor $\nu= N_{S}\Phi_{0}/B$ ($N_{S}$ and
$\Phi_{0}$ being the two-dimensional carrier concentration and the
flux quantum respectively). These studies have been extended to
the case of non-integer values of $\nu$ \cite{By1} and have also
included the calculation of matrix elements for  the optical
conductivity \cite{By2}. The effects of EEI  in Graphene have
recently been reported on a theoretical basis \cite{Iyengar} but
with a different model than that of Refs. \cite{KH,KM,By1} and
restricted to integer values of the filling factor. We have
followed here the lines of Ref. \cite{By2}, which has been shown
to reproduce \textbf{quantitatively} experimental results
\cite{Faugeras} when they are interpreted in terms of MP
excitations.

Because of the Kohn's theorem \cite{Kohn}, the EEI effects turn
out to be tiny for C2DEG. However this theorem does not apply for
a linear dispersion law and therefore EEI are expected to induce
significant effects in Graphene. Indeed  recent experimental
investigations of the magneto-optical transitions in Graphene
\cite{Sadowski,Jiang} have been interpreted with an effective
velocity $\widetilde{c}$, replacing $v_{F}$ in Eq.1, ranging
between $ 1.03$ to $1.18 \times 10^{6}m/s$ and showing a
re-normalization  of $v_{F}$ which here, we will show, is mainly
due to electron-electron interactions.

On general grounds, the  Coulomb energy characteristic of
electron-electron interaction in magnetic field is
$E_{c}=e^{2}/\kappa l_{B}$ where $\kappa$ is the electronic
dielectric constant of the material and $l_{B}= (eB)^{-1/2}$ the
magnetic field length. The magnetoplasmon  approach assumes that
$E_{c}$ is smaller than the one-electron energy transitions. In
the present case of Graphene $E_{c}(meV)=11.2\sqrt{B(T)}$ and
$E_{10}(meV)= 31\sqrt{B(T)}$ (see Eq.1) leading to the ratio
$E_{10}/E_{c}= 2.77$ that is a condition better fulfilled for
Graphene than for GaAs based C2DEG and furthermore not dependent
on the value of the magnetic field. Magneto-excitons should be
therefore more stable in Graphene than in C2DEG and the approach
derived for these later compounds should be valid. The report is
organized as follows: we will first describe the general formalism
used to derive the MP dispersion curves (section II). We apply it
to the case of filling factors $\nu<2$ in section III and to the
case of  $2<\nu<6$ in section IV. Results will be discussed and
compared to experimental results in section V. The details of
calculations are reported in the appendices.

\section{ General formalism}

In contrast to the GaAs case, Graphene has two valleys which lead
to the conclusion that in the absence of spin splitting and valley
splitting each Landau Level (LL) is, in general,  four times
degenerate. The fourfold degeneracy of the  $n=0$ LL is still due
to spin and valley symmetries, but two of these levels have an
electron-like character and the two other ones a hole-like
character.  We will restrict our analysis to the zero temperature
case. Because of the peculiar symmetry of the problem the wave
functions have a spinor character which can be expressed in the
Landau gauge, with the potential vector components of the magnetic
field $A_{x}=A_{z}=0$ and $A_{y}= Bx$, as:

\begin{align}
F_{np}^{K}(\overrightarrow{\rho})&=\frac{c_{n}}{\sqrt L}e^{\imath
py}{-\imath sgn(n)\varphi_{|n|-1}(x-p)\choose
\varphi_{|n|}(x-p)}\notag\\
F_{np}^{K'}(\overrightarrow{\rho})&=\frac{c_{n}}{\sqrt L}e^{\imath
py}{\varphi_{|n|}(x-p)\choose -\imath sgn(n)\varphi_{|n|-1}(x-p)}
\end{align}
where $\overrightarrow{\rho}$ is the two-dimensional vector of
components $x$ and $y$, $c_{n}=1$ for $n=0$ and $1/\sqrt 2$
otherwise whereas $sign(n)= 1,0,-1$ when $n>0,=0,<0$ respectively.
 $ \varphi_{|n|}(x)$ is the standard normalized Landau wave
functions. Note that these wave functions differ from those
proposed in Ref. \cite{Zheng} by the phase factor due to the
different gauge used. Following the lines of Ref. \cite{By1}, we
call ${\cal{A}}_{n,n',\sigma,i}^{+}$ the creation operator of an
exciton of energy $E_{ex}(\vec{k})$, corresponding to a transition
from LL $n'$ with spin $\sigma$ in valley $i$ to a LL $n$ of the
same spin and same valley. This operator is defined as a function
of  $a_{\lambda}^{+}$  and $a_{\lambda}$,  the standard one
particle creation and annihilation operators  respectively, as:

\begin{equation}
\begin{split}
{\cal{A}}_{n,n',\sigma,i}^{+}(\vec{k})|0\rangle &=
   \sum_{p}\exp (ik_{x}(p+k_{y}/2))\\
   &\times a_{n,p,\sigma,i}^{+}a_{n',p+k_{y},\sigma,i}|0\rangle
\end{split}
\end{equation}

The total Hamiltonian of the system is written as:
\begin{equation}
\widehat{H}_{tot}=
\sum_{m,p,\sigma,i}\hbar\omega_{m}^{\sigma,i}a_{m,p,\sigma,i}^{+}a_{m,p,\sigma,i}
    +\widehat{H}_{int}
\end{equation}

where $\hbar\omega_{m}^{\sigma,i}$ is the corresponding one
electron energy of the LL $m$ with spin $\sigma$ in valley $i$.
The Coulomb interactions appear in $\widehat{H}_{int}$ as :

\begin{equation}
\begin{split}
    \widehat{H}_{int}&= \frac{1}{2}\int
    d\vec{\rho}_{1}d\vec{\rho}_{2}V(\vec{\rho}_{1}-\vec{\rho}_{2})\\
    &\times[\widehat{F}_{\sigma_{1},i_{1}}^{+}(\vec{\rho}_{1})
    [\widehat{F}_{\sigma_{2},i_{2}}^{+}(\vec{\rho}_{2})
    \widehat{F}_{\sigma_{2},i_{2}}(\vec{\rho}_{2})]
    \widehat{F}_{\sigma_{1,i_{1}}}(\vec{\rho}_{1})]
\end{split}
\end{equation}

where $[\widehat{F}^{+}\widehat{F}]$ denotes the scalar product
and

\begin{align}
\widehat{F_{\sigma,i}}(\overrightarrow{\rho})&=
\sum_{n,p}F_{n,p}^{i}(\overrightarrow{\rho})
a_{n,p,\sigma,i}\nonumber\\
V(\overrightarrow{\rho_{1}}-\overrightarrow{\rho_{2}})&=
\int\frac{d^{2}q}{(2\pi)^{2}}\widetilde{V(q)}e^{\imath
\overrightarrow{q}\cdot(\overrightarrow{\rho_{1}}-\overrightarrow{\rho_{2}})}.
\end{align}

 with $\widetilde{V}(q)$ being the
 2D Fourier transform of the Coulomb potential $V(r)=e^{2}/(\kappa r)$.

After some calculations we obtain an analytic expression for
$\widehat{H}_{int}$ which is expressed as:

\begin{eqnarray}
\widehat{H}_{int}= \frac{1}{2}\sum
\widetilde{V}(q)\exp(iq_{x}(p_{1}-p_{2}-q_{y}))\times \nonumber\\
 \widetilde{J_{n_{4},n_{1}}(\vec{q})}\widetilde{J_{n_{3},n_{2}}(-\vec{q})}\times \nonumber\\
 a_{n_{1},p_{1},\sigma_{1},i_{1}}^{+}a_{n_{2},p_{2},\sigma_{2},i_{2}}^{+}
 a_{n_{3},p_{2}+q_{y},\sigma_{2},i_{2}}a_{n_{4},p_{1}-q_{y},\sigma_{1},i_{1}}
\end{eqnarray}

In Eq.7, the summation is extended over the ensemble
$n_{1},n_{2},n_{3},n_{4}$ of LL, the ensemble
 $p_{1},p_{2}$ of the y-component of the momentum,
  the ensemble of spin $\sigma_{1},\sigma_{2}$, both valleys
  $i_{1}$ and $i_{2}$ and the wavevector $\vec{q}$.

The function $\widetilde{J_{m,n}}$ is defined as:
\begin{equation}
\begin{split}
\widetilde{J_{m,n}}(\vec{q})=
c_{n}^{*}c_{m}&\{sgn(m)sgn(n)J_{|m|-1,|n|-1}(\vec{q})\\
 &+ J_{|m|,|n|}(\vec{q})\}
\end{split}
\end{equation}

with the usual definition of the integral $J_{m,n}(\vec{q})$ valid
for $m>n$:

\begin{align}
J_{m,n}(\vec{q})&=\int dx e^{\imath
q_{x}x}\varphi_{m}(x+\frac{q_{y}}{2})\varphi_{n}(x-\frac{q_{y}}{2})\nonumber\\
&=(\frac{n!}{m!})^{\frac{^1}{2}}e^{-\frac{q^{2}}{4}}(\frac{q_{y}+iq_{x}}{\sqrt{2}})^{m-n}
 L_{n}^{m-n}(\frac{q^{2}}{2})
\end{align}

where $L_{n}^{m-n}(x)$ are the Laguerre polynomials. For $m<n$ the
relation $J_{m,n}(\vec{q})=J_{n,m}^{*}(-\vec{q})$ holds.

Using the random phase approximation (RPA) to treat the
combination of creation and annihilation operators we arrive to
the following expression for the Exciton energies (the notation
 $|0\rangle$ representing the ground state of the system) :

\begin{eqnarray}
&&
E_{ex}(\vec{k}){\cal{A}}_{n,n',\sigma,i}^{+}|0>=\hbar\omega_{n,n',i}^{\sigma}{\cal{A}}_{n,n',\sigma,i}^{+}(\vec{k})|0\rangle
\nonumber\\
&&\hspace{-0.5cm}+\sum_{n_{2}}[\widetilde{E}_{n',n_{2},n',n_{2}}(0)-\widetilde{E}_{n,n_{2},n,n_{2}}(0)]
f_{n_{2},i}^{\sigma}\nonumber\\
&&\hspace{3cm}\times{\cal{A}}_{n,n^{'},\sigma,i}^{+}(\vec{k})|0\rangle \nonumber\\
&&\hspace{-0.5cm}+\sum_{n_{2},n_{4}}\widetilde{E}_{n',n_{2},n,n_{4}}(k_{y},k_{x})(f_{n,i}^{\sigma}-f_{n',i}^{\sigma}){\cal{A}}_{n_{2},n_{4},\sigma,i}^{+}
(\vec{k})|0\rangle\nonumber\\
&&\hspace{-0.5cm}-\sum_{n_{2},n_{3},\sigma_{2},j}\frac{\widetilde{\widetilde{V}}_{n',n_{2},n_{3},n}(-k_{x},k_{y})}{2\pi}(f_{n,i}^{\sigma}-f_{n',i}^{\sigma})\nonumber\\
&&\hspace{3cm}\times{\cal{A}}_{n_{2},n_{3},\sigma_{2},j}^{+}(\vec{k})|0\rangle \nonumber\\
\end{eqnarray}

where  $f_{n,i}^{\sigma}$ is the filling factor of LL $n$ with
spin $\sigma$ in valley $i$. The matrix elements $\widetilde{E}$
and $\widetilde{\widetilde{V}}$ are given by:
\begin{align}
\widetilde{\widetilde{V}}_{n_{1},n_{2},n_{3},n_{4}}(\vec{q})&=\widetilde{V}(q)\widetilde{J_{n_{4},n_{1}}}(\vec{q})
\widetilde{J_{n_{3},n_{2}}}(-\vec{q})\nonumber\\
\widetilde{E}_{n_{1},n_{2},n_{3},n_{4}}(\vec{k})&=\int\frac{d\vec{q}}{(2\pi)^{2}}\widetilde{\widetilde{V}}_{n_{1},n_{2},n_{3},n_{4}}(\vec{q})
e^{i\vec{k}\cdot\vec{q}}
\end{align}

We note, at that level, that Eq. 10 is formally equivalent to that
obtained for C2DEG \cite{By1} except for the definition of the
different matrix elements which here takes into account the spinor
character of the wave functions. In this equation, the second term
(second line) is a measure of the difference of exchange energies
of the LL $n'$ and $n$. The third line is related to the direct
electron-hole Coulomb interaction (Exciton binding energy). Both
terms are involving excitons of same spin and same valley. The
last term of Eq. 10 describes the simultaneous annihilation and
creation of excitons at different points of the Brillouin zone
(RPA contribution): it includes  all possible transitions without
restriction to spin or valley indices. The exchange terms deserve
a special attention in the present case. The corresponding
expression for the exchange in Eq. 10 reads as:

\begin{multline}
\hspace{-0.3cm}\widetilde{E}_{n,m,n,m}(0)=|c_{n}|^{2}|c_{m}|^{2}\sqrt{2}\int
dx
e^{-x^{2}}\\
\{L_{|m|})L_{|n|}
+ (sgn(n)sgn(m))^{2}L_{|m|-1}L_{|n|-1}\\
 + sgn(n)sgn(m)\frac{2x^{2}}{\sqrt{|m||n|}}L_{|m|-1}^{1}L_{|n|-1}^{1}\}
\end{multline}
where all Laguerre polynomials have arguments $x^{2}$.
 In Eq.10
(second line), the summation over $n_{2}$, for these terms, has to
include all LL from $-\infty$ to $+\infty$. The evaluation of the
exchange contributions to the different situations is given in
appendix A.

Solving Eq.10 results in diagonalizing an Hamiltonian, the size of
which depends on the number of transitions which are assumed to be
coupled by electron-electron interactions. In reality this number
is very large for Graphene due to the existence of interband
transitions but since $E_{c}$ is smaller than the energy of
transitions, we can reasonably assume, in the spirit of the HF
approximation, that EEI will not couple, at first order,
transitions with different one-electron energies. In that case,
the problem reduces to solve the Hamiltonian for each type of
optical transitions which depend on the value of the filling
factor. However, even if one solves the whole problem in
successive steps, one has to keep in mind that  a \textbf{common}
energy scale should be adopted for all transitions in order to
compare such results with experimental ones.

When writing the Hamiltonian, using Eq. 10, for a given set of
transitions in the basis $\overrightarrow{\phi}$ corresponding to
these transitions such that
$\widehat{H}_{tot}\overrightarrow{\phi}=
E_{ex}(\overrightarrow{k})\overrightarrow{\phi}$, we end up with a
matrix which is not symmetric, as in C2DEG \cite{By1}, but in
addition, here, many matrix elements are complex as this will be
seen in  appendix B. To make the treatment easier to follow, we
adopt the same technique as used in \cite{By1} which consists in
writing the hamiltonian in a new basis $\overrightarrow{\Psi}=
\widehat{M}\overrightarrow{\phi}$ where $\widehat{M}$ is a
diagonal unitary matrix. The new Hamiltonian is then expressed as:
\begin{equation}
\widetilde{H}=\widehat{M}\widehat{H}_{tot}\widehat{M}^{-1}
\end{equation}
 which is now symmetric and has only real matrix elements.

In the calculations we will neglect the spin splitting
$\Delta_{S}$ which is small in the case of Graphene \cite{Zhang1}
but will not change anyway the conclusions since the optical
transitions conserve the spin.

We will furthermore assume in the following that there exists some
valley splitting $\Delta_{V}$ first suggested by Gusynin
\textit{et al.} \cite{Gusynin}. The existence of such a valley
splitting,  has been recently supported by different models. Some
of these models include different types of electron-phonon
interactions \cite{Fuchs,Yan}: they all predict a linear
dependence of $\Delta_{V}$ with the magnetic field. Another one
\cite{Abanin} invokes EEI with strain induced gauge field yielding
to a valley splitting which varies like $\sqrt{B}$. We assume
here, for convenience, that $\Delta_{V}$ is larger than
$\Delta_{S}$  in such a way the electrons remain in the same
valley ( here the valley \textbf{K} for instance) for filling
factor $\nu<1$. This is not necessary true since experimental
results \cite{Jiang1} tend to favor a situation, where for $\nu
\ll 1$, the system becomes spin polarized. The splitting
$\Delta_{V}$ will not be included in the present calculations but
its consequence will be discussed in each case where its
contribution could be relevant. All the following results for
energies are given in units of $E_{c}$ and as a function of
$K=|\overrightarrow{k}l_{B}|$.

\section{ Magnetoplasmon energies for $\nu < 2$}

\begin{figure}[h]
\centering
\includegraphics[width=1.0\columnwidth]{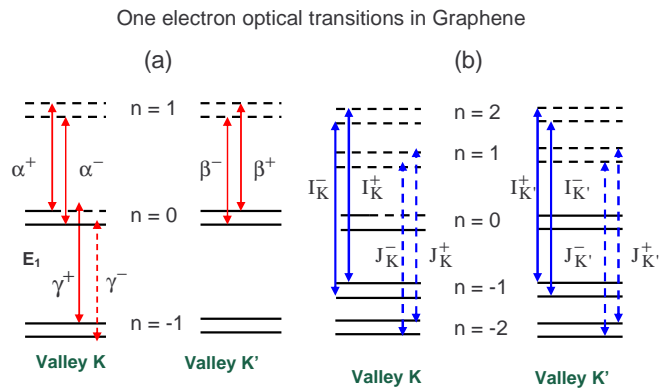}
\caption{\label{F1}(Color on line) Schematic diagram of
one-electron transitions used in the magnetoplasmon model for $
1<\nu<2$. On the left part of the figure (a) are shown the
transitions implying the $n=0$ LL. On the right part of the figure
(b) are shown the first interband (electron-hole) transitions from
the $n=-2,-1$ LL to the $n=1,2$ LL respectively. For $\nu<1$, the
transition $\alpha^{+}$ in (a) disappears and the new transition
$\gamma^{-}$ appears.The splitting of LL $n$ mimics the spin
splitting.}
\end{figure}

 For $\nu<2$, the typical transitions to be analyzed are
  displayed in Fig.1 . There are two kinds of one
electron transitions, in each valley, those which imply the $n=0$
LL (Fig.1a) and those which correspond to interband
(electron-hole) transitions (in Fig.1b are only represented those
from LL $n =-2,-1$ to $n = 1,2$).
  As already mentioned in the preceding section
  one can treat independently the one-electron
transitions implying the $n=0$ LL  and those involving the
interband transitions.

\subsection{ Magnetoplasmon energies for transitions implying
the $n=0$ LL}

The different one-electron energy transitions which are considered
in this case are displayed in Fig.1a for $1<\nu<2$. In this figure
the splitting of LL $n$ mimics the spin-splitting $\Delta_{S}$ for
clarity but, as already said, this splitting is not taken into
account in the present calculations. The Hamiltonian to be solved
is therefore a matrix of rank 5 written, first, in the basis
$\overrightarrow{\phi}=(\alpha^{+},\alpha^{-},\beta^{+},\beta^{-},\gamma^{+})$
(see Fig. 1a for notations) and then transformed according to Eq.
13. The corresponding diagonal matrix $\widehat{M}$ is denoted
here $\widehat{M0}_{1<\nu<2}$ and has the following diagonal
elements:

 \begin{equation}
\widehat{M0}_{1<\nu<2}=\{\frac{e^{-\imath\varphi}}{\sqrt{f_{0}^{+}}},e^{-\imath\varphi},e^{-\imath\varphi},
e^{-\imath\varphi},\frac{e^{\imath\varphi}}{\sqrt{1-f_{0}^{+}}}\}
\end{equation}
where $f_{0}^{+}$ is the partial filling factor of the spin-up
$n=0$ LL and $\varphi$  the polar angle of the exciton wave
vector.  For $1<\nu<2$, the matrix $\widetilde{H_{1<\nu<2}^{0}}$
 to be diagonalised is expressed as:

\begin{widetext}
\begin{equation}
\widetilde{H_{1<\nu<2}^{0}}=
\begin{bmatrix}h_{11}&\sqrt{f_{0}^{+}}V_{0101}&\sqrt{f_{0}^{+}}V_{0101}&\sqrt{f_{0}^{+}}V_{0101}
&\sqrt{f_{0}^{+}(1-f_{0}^{+})}(V_{0011}-E_{0011})\\
\sqrt{f_{0}^{+}}V_{0101}&h_{22}&V_{0101}&V_{0101}&\sqrt{1-f_{0}^{+}}V_{0011}\\
 \sqrt{f_{0}^{+}}V_{0101}&V_{0101}&h_{33}&V_{0101}&\sqrt{1-f_{0}^{+}}V_{0011}\\
\sqrt{f_{0}^{+}}V_{0101}&V_{0101}&V_{0101}&h_{44}&\sqrt{1-f_{0}^{+}}V_{0011}\\
\sqrt{f_{0}^{+}(1-f_{0}^{+})}(V_{0011}-E_{0011})&\sqrt{1-f_{0}^{+}}V_{0011}&\sqrt{1-f_{0}^{+}}V_{0011}
&\sqrt{1-f_{0}^{+}}V_{0011}&h_{55}\end{bmatrix}
\end{equation}
\end{widetext}
The different matrix elements are given in appendices A  (Eq. A1)
and B (Eqs .B1, B2, B3).

One can note that the eigenvalues of $\widetilde{H_{1<\nu<2}^{0}}$
are identical for $f_{0}^{+}$= 0 or 1 (that is $\nu$=1 or 2)
whereas those for non integer values of $\nu$ are symmetric with
respect to $\nu=1.5$. For $\nu<1$ the corresponding Hamiltonian
$\widetilde{H_{0<\nu<1}^{0}}$ has to be written in the basis
$\overrightarrow{\phi}=(\alpha^{-},\beta^{+},\beta^{-},\gamma^{+},\gamma^{-})$
(see Fig1a), replacing $f_{0}^{+}$ by $f_{0}^{-}$ with a new
diagonal matrix $\widehat{M}$ denoted now $\widehat{M0}_{0<\nu<1}$
which has the following  elements:

 \begin{equation}
\widehat{M0}_{0<\nu<1}=\{\frac{e^{-\imath\varphi}}{\sqrt{f_{0}^{-}}},e^{-\imath\varphi},e^{-\imath\varphi},
e^{\imath\varphi},\frac{e^{\imath\varphi}}{\sqrt{1-f_{0}^{-}}}\}
\end{equation}

where $f_{0}^{-}$ is the partial filling factor of the spin-down
$n=0$ LL.

The corresponding expressions for the matrix elements are given in
appendices A  (Eq. A1) and B (Eqs .B1, B2, B3). It turns out that
the eigen values of $\widetilde{H_{0<\nu<1}^{0}}$ are symmetric of
those obtained for $\widetilde{H_{1<\nu<2}^{0}}$ with respect to
$\nu=1$. If we adopt a model where  $\Delta_{V}$ is smaller than
$\Delta_{S}$ for $\nu<1$, we obtain an Hamitonian with the same
eigen values which shows that the MP results do not depend on this
assumption.

Results for the MP dispersion curves are displayed in Fig.2.  For
$\nu=1$ or $2$ one obtains, for the dispersion curves, a solution
$Ed(K)$ three times degenerate and one solution $Eu(K)$ which have
the following analytical expressions:
\begin{align}
Ed(K)&= E_{10}+ C_{1} +\frac{3}{4}\alpha_{0}-E_{0110}(K)\nonumber\\
Eu(K)&= Ed(K)+ 4V_{0101}(K)
\end{align}
where  $E_{10}= E_{1}-E_{0}= 2.77 \times e^{2}/(\kappa l_{B})$ is
the one-electron energy for these transitions and $C_{1}$ defined
in appendix A (Eq. A2) is a quantity a priori divergent which will
be discussed in section V.

 For non integer values of $\nu$, the solutions $Ed(K)$
remain twice degenerate, the high energy solution remains close to
$Eu(K)$ and two new solutions appear. The linear dispersion near
$K\simeq 0$ for $Eu(K)$ is due to the RPA contribution entering
Eq. 10. As compared to the solutions found in C2DEG for $\nu=2$
\cite{KH,By2}, this contribution is the same whereas that of the
exciton binding energy is different. The solutions for $K\simeq 0$
will be further discussed in section V below.

\begin{figure}[ht]
\centering
\includegraphics[width=1.0\columnwidth]{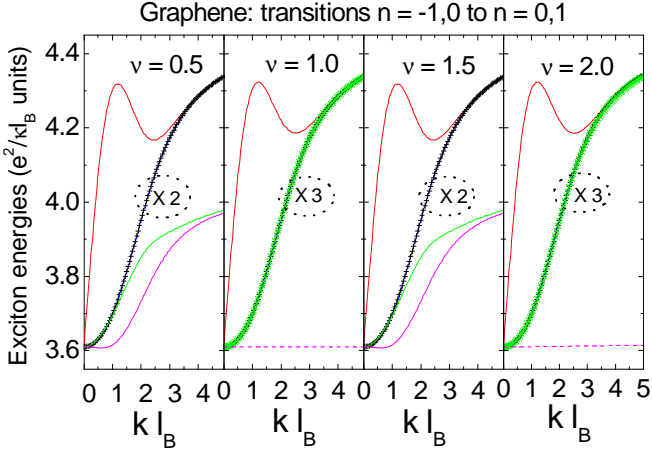}
\caption{\label{F2} (Color on line) Variation of the
magnetoplasmon energies in units of $e^{2}/(\kappa l_{B})$ as a
function of $kl_{B}$ for the transitions involving the $n=0$ LL
and filling factors $\nu <2$. The dotted circles denote the
degeneracy of the transitions.}
\end{figure}

Following the lines of Ref.\cite{By2} we have also calculated, in
the frame of the MP picture, the optical conductivity (see
appendix C, Eq. C4) which predicts that $Eu(K)$ should be
optically active in both polarizations of the light (note that the
optical vectors are proportional to $v_{F}^{2}$).

The MP model has been derived without including the  valley
splitting $\Delta_{V}$: if such a splitting is introduced, for the
$n=0$ LL, we expect  a corresponding splitting of the optical
transition independent on the relative magnitude of $\Delta_{V}$
and $\Delta_{S}$.

\subsection{ Magnetoplasmon energies for transitions from the
$n=-2,-1$ to $n=1,2$ LL}

We discuss now the case of interband transitions displayed in
Fig.1b. There are, in this case, eight possible  one-electron
transitions and the Hamiltonian is written first in the basis
$\overrightarrow{\phi}=(I_{K}^{-},J_{K}^{-},I_{K}^{+},J_{K}^{+},I_{K'}^{-},J_{K'}^{-},I_{K'}^{+},J_{K'}^{+})$
and then transformed according to Eq.13.
 For $0<\nu<2$, the corresponding  diagonal matrix $\widehat{M}$ denoted here as $\widehat{MI}_{0<\nu<2}$ has the following
diagonal elements:
\begin{equation}
\widehat{MI}_{0<\nu<2}=
\{e^{-\imath\varphi},e^{\imath\varphi},e^{-\imath\varphi},
e^{\imath\varphi},e^{-\imath\varphi},e^{\imath\varphi},e^{-\imath\varphi},e^{\imath\varphi}\}
\end{equation}

For these transitions the symmetrized excitonic Hamiltonians
$\widetilde{HI^{12}_{1<\nu<2}}$ and
$\widetilde{HI^{12}_{0<\nu<1}}$ have matrix elements which are
given in appendices A (Eqs. A3, A4) and B (Eqs. B4, B5, B6).

 The dispersion of MP energies, in units of $E_{c}$, are displayed in
Fig.3 as a function of $K$. The corresponding one electron energy
for these transitions is $EI_{12}=E_{1}-E_{-2}=(\sqrt{2}+1)E_{10}=
6.69 \times e^{2}/(\kappa l_{B})$.

\begin{figure}
\centering
\includegraphics[width=0.9\columnwidth]{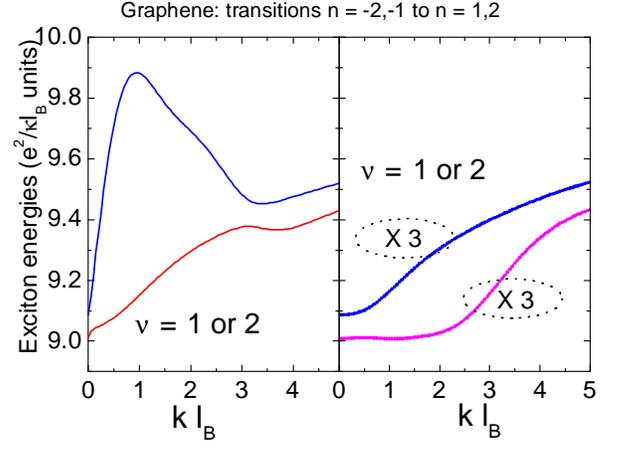}
\caption{\label{F3}(Color on line) Variation of the magnetoplasmon
energies in units of $e^{2}/(\kappa l_{B})$ as a function of
$kl_{B}$ for the transitions $n=-2,-1$ to $n=1,2$ LL and filling
factor $\nu =1$ or $2$.  The dotted circles denote the degeneracy
of the transitions.The magnetoplasmon curves in the left panel are
not degenerate and are the only optically active transitions.}
\end{figure}

As for the preceding case, the solutions are identical for $\nu=1$
or $2$ and symmetric with respect to $\nu=1$. For integer values
of $\nu$, the eigen-values of the Hamiltonian can be expressed
analytically and arranged in two groups: (i) two single solutions
$EI_{1}^{+/-} (K)$ displayed in the left part of the Fig.3 and
(ii) two other sets of solutions $EI_{2}^{+/-}(K)$, three times
degenerate (see dotted circles in Fig.3),  displayed in the right
part of the figure. They are expressed as:

\begin{figure}
\centering
\includegraphics[width=0.9\columnwidth]{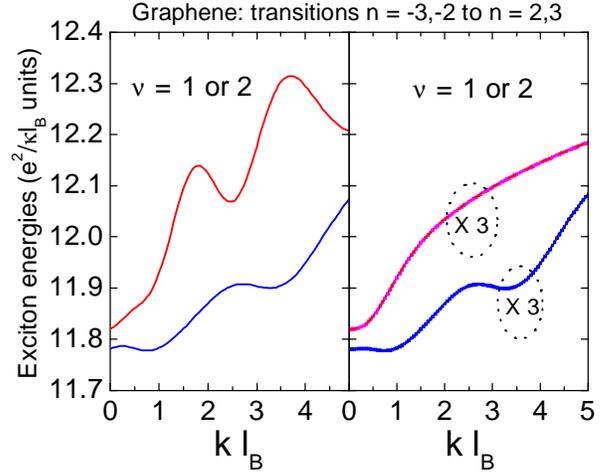}
\caption{\label{F4}(Color on line) Variation of the magnetoplasmon
energies in units of $e^{2}/(\kappa l_{B})$ as a function of
$kl_{B}$ for the transitions $n=-3,-2$ to $n=2,3$ LL and filling
factor $\nu =1$ or $2$.  The dotted circles denote the degeneracy
of the transitions.The magnetoplasmon curves in the left panel are
not degenerate and are the only optically active transitions.}
\end{figure}

\begin{equation}
\begin{split}
EI_{1}^{+/-}(K)=&(\sqrt{2}+1)(E_{10}+C_{1})+ \Delta C_{2}\\
&+ 4V_{-12-12} -E_{-122-1}\\
&\pm\sqrt{(\frac{\alpha_{0}}{16})^{2}+
(4V_{-12-12}-E_{-112-2})^{2}}
 \end{split}
\end{equation}

 and

\begin{equation}
\begin{split}
EI_{2}^{+/-}(K)=(\sqrt{2}+1)(E_{10}+C_{1})+ \Delta C_{2}\\
-E_{-122-1} \pm \sqrt{(\frac{\alpha_{0}}{16})^{2}+
(E_{-112-2})^{2}}
\end{split}
\end{equation}
where all matrix elements entering Eqs. 19 and 20 are function of
$K$ and given in appendices A (Eq. A4) and B (Eqs. B5, B6).
 Only the solutions $EI_{1}^{+/-}(K)$ are optically active (see
appendix C, Eq. C5). For non-integer values of the filling factor
the results are very close to those presented in Fig.3 except for
two solutions of the two groups of degenerate transitions which
are no longer degenerate for $K\simeq 0$.

In contrast to the case of transitions implying the $n=0$ LL, a
splitting of the transitions equal to $\alpha_{0}/8$ and due to
electron-electron interactions is expected for $K\simeq 0$.

Note however, here, that the introduction of a valley splitting
$\Delta_{V}$ should only provide an additional component either
linear in $B$ or in $\sqrt{B}$ depending on the origin of this
valley splitting.

\subsection{ Magnetoplasmon energies for transitions from the
$n=-3,-2$ to $n=2,3$ LL}

In this case  the corresponding Hamiltonian
$\widetilde{HI^{23}_{0<\nu<2}}$
 has the same structure that $\widetilde{HI^{12}_{0<\nu<2}}$ and therefore only
 the values of matrix elements are different. They are given in
appendices A (Eqs. A5, A6) and B (Eqs. B7, B8). The dispersion of
MP energies, in units of $E_{c}$, is displayed in Fig.4 as a
function of $k l_{B}$. The corresponding one electron energy for
these transitions is $EI_{23}=E_{2}-E_{-32}=(\sqrt{3}+
\sqrt{2})E_{10}= 8.72 \times e^{2}/(\kappa l_{B})$. The solutions
are formally identical to those given in Eqs. 19 and 20 with the
appropriate changes for the matrix elements  given in the
appendices A (Eq. A5) and B (Eqs. B7, B8). The splitting of the
transitions for $K\simeq 0$ is here equal to $ \alpha_{0}/16$.

We, now, evaluate the exciton energies for $\nu> 2$.

\section{ Magnetoplasmon energies for $2< \nu <6$}

We will  concentrate the report for filling factors $2<\nu<6$. The
contributions of exchange are given in appendix A (Eq. A7). It
turns out that the problem to solve is symmetric with respect to
$\nu =4$ and therefore we will detail the treatment for $2<\nu<4$
and will note only the main changes for $4<\nu<6$.

\subsection{ Magnetoplasmon energies for $2<\nu<4$}

In this case we have to treat the problem depicted in Fig.5 for
the one electron energy transitions. Note that we have here two
types of transitions those implying the $n=0$ LL and those between
LL $n=1$ and $n=2$. Because the corresponding one electron
energies are different they are treated independently.
\linebreak[2]

\begin{figure}[h]
\centering
\includegraphics[width=0.9\columnwidth]{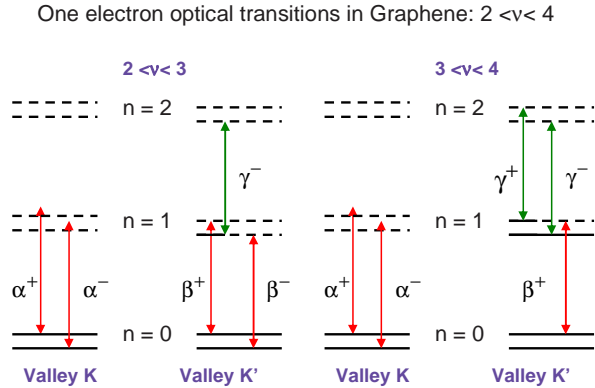}
\caption{\label{F5}(Color on line) Schematic diagram of
one-electron transitions used in the magnetoplasmon model for $
2<\nu<4$.}
\end{figure}

For $2<\nu<3$, we have to write, first, the Hamiltonian in the
basis
$\overrightarrow{\phi}=\{\alpha^{+},\alpha^{-},\beta^{+},\beta^{-},\gamma^{-}\}$
and for $3<\nu<4$ in the basis
$\overrightarrow{\phi}=\{\alpha^{+},\alpha^{-},\beta^{+},\gamma^{-},\gamma^{+}\}$
and transformed them according to Eq. 13. For $2<\nu<3$ and
$3<\nu<4$, the corresponding  diagonal matrices $\widehat{M}$,
denoted here as $\widehat{M12}_{2<\nu<3}$ and
$\widehat{M12}_{3<\nu<4}$ respectively, have the following
diagonal elements:

\begin{align}
\widehat{M12}_{2<\nu<3}=\{1,1,1,\frac{1}{\sqrt{1-f_{1}^{-}}}
,\frac{1}{\sqrt{f_{1}^{-}}}\}\nonumber\\
\widehat{M12}_{3<\nu<4}=\{1,1,\frac{1}{\sqrt{1-f_{1}^{+}}},
1,\frac{1}{\sqrt{f_{1}^{+}}}\}
\end{align}
where $f_{1}^{-}$ and $f_{1}^{+}$ are the partial filling factors
of the spin-down and spin-up  $n=1$ LL respectively, attached to
the valley K' with our convention. One gets for the corresponding
Hamiltonians $\widetilde{H_{2<\nu<3}^{12}}$ and
$\widetilde{H_{3<\nu<4}^{12}}$ the following expressions:

\begin{widetext}
\begin{equation}
\widetilde{H_{2<\nu<3}^{12}}=
\begin{bmatrix}h_{11}&V_{0101}&V_{0101}&\sqrt{1-f_{1}^{-}}V_{0101}&0\\
V_{0101}&h_{22}&V_{0101}&\sqrt{1-f_{1}^{-}}V_{0101}&0\\
 V_{0101}&V_{0101}&h_{33}&\sqrt{1-f_{1}^{-}}V_{0101}&0\\
\sqrt{1-f_{1}^{-}}V_{0101}&\sqrt{1-f_{1}^{-}}V_{0101}&\sqrt{1-f_{1}^{-}}V_{0101}&h_{44}&0\\
0&0&0&0&h_{55}\end{bmatrix}
\end{equation}
and
\begin{equation}
\widetilde{H_{3<\nu<4}^{12}}=
\begin{bmatrix}h_{11}&V_{0101}&\sqrt{1-f_{1}^{+}}V_{0101}&0&0\\
V_{0101}&h_{22}&\sqrt{1-f_{1}^{+}}V_{0101}&0&0\\
 \sqrt{1-f_{1}^{+}}V_{0101}&\sqrt{1-f_{1}^{+}}V_{0101}&h_{33}&0&0\\
0&0&0&h_{44}&\sqrt{f_{1}^{+}}V_{1212}\\
0&0&0&\sqrt{f_{1}^{+}}V_{1212}&h_{55}\end{bmatrix}
\end{equation}
\end{widetext}

where the new matrix elements entering these matrices are given in
appendices A (Eqs. A7) and B (Eqs. B9, B10). The resulting
dispersion curves are displayed in Fig.6. The corresponding
one-electron energies for both types of transitions are $E_{10}=
2.77 \times e^{2}/(\kappa l_{B})$ for the higher ones and
$E_{12}=E_{2}-E_{1}= 1.15 \times e^{2}/(\kappa l_{B})$ for the
lower ones.

\begin{figure}[ht]
\centering
\includegraphics[width=1.0\columnwidth]{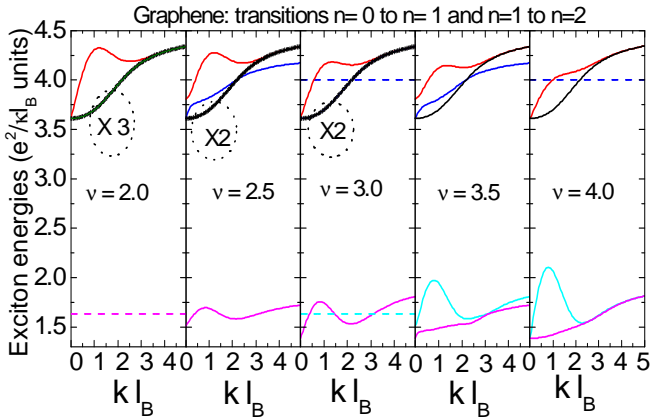}
\caption{\label{F6} (Color on line) Variation of the
magnetoplasmon energies in units of $e^{2}/(\kappa l_{B})$ as a
function of $kl_{B}$ for different filling factors $2<\nu <4$.The
dotted circles denote the degeneracy of the transitions. }
\end{figure}

The eigen-values of the Hamiltonians (Eqs. 22 and 23) can be
expressed analytically. For $2<\nu<3$ there are two solutions
identical to $Ed(K)$ (Eq.17) which remain  degenerate and which
are not optical active. Two other non degenerate solutions denoted
as $E0^{\pm}_{2<\nu<3}(K)$ are given by:

\begin{equation}
\begin{split}
E0_{2<\nu<3}^{\pm}(K)=\frac{1}{2}[h_{11}+h_{44}+2V_{0101}\\
\pm\sqrt{(h_{44}-h_{11}-2V_{0101})^{2}+12
V_{0101}^{2}(1-f_{1}^{-})}]
\end{split}
\end{equation}
and  a third one $E12_{2<\nu<3}(K)= h_{55}(K)$.  These transitions
are all optically active with a relative weight depending on the
filling factor.

For $3<\nu<4$, it remains one eigen-value solution identical to
$Ed(K)$ (Eq.17) and two groups of optically active non degenerate
solutions denoted as $E0_{3<\nu<4}^{\pm}(K)$ and
$E1_{3<\nu<4}^{\pm}(K)$ with the following analytical expressions:

\begin{equation}
\begin{split}
E0_{3<\nu<4}^{\pm}(K)=\frac{1}{2}[h_{11}+h_{33}+2V_{0101}\\
\pm\sqrt{(h_{33}-h_{11}-2V_{0101})^{2}+ 8
V_{0101}^{2}(1-f_{1}^{+})}]
\end{split}
\end{equation}
and:

\begin{equation}
\begin{split}
E1_{3<\nu<4}^{\pm}=\frac{1}{2}[h_{44}+h_{55}\\
\pm\sqrt{(h_{44}-h_{55})^{2}+ 4 V_{1212}^{2}f_{1}^{+}}]
\end{split}
\end{equation}
for which all matrix elements are function of $K$.

The corresponding optical vectors for these transitions are given
in appendix C (Eqs. C7 and C8).
\begin{figure}[ht]
\centering
\includegraphics[width=1.0\columnwidth]{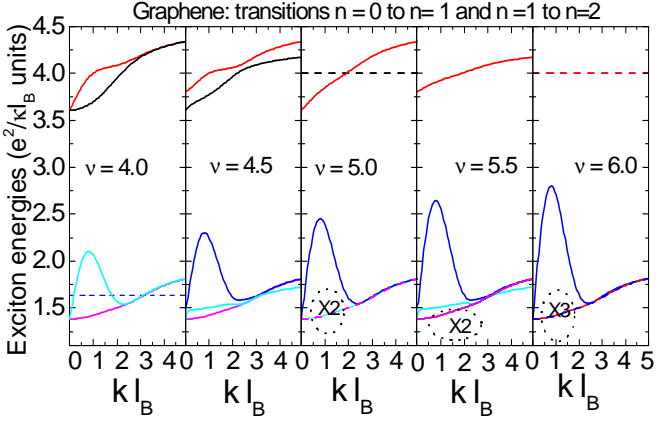}
\caption{\label{F7} (Color on line) Variation of the
magnetoplasmon energies in units of $e^{2}/(\kappa l_{B})$ as a
function of $kl_{B}$ for different filling factors $4<\nu <6$. The
dotted circles denote the degeneracy of the transitions. }
\end{figure}

\subsection{ Magnetoplasmon energies for $4<\nu<6$}

It is easy to see that, in this case, the structures of the
corresponding Hamiltonians $\widetilde{H_{4<\nu<5}^{12}}$ and
$\widetilde{H_{5<\nu<6}^{12}}$ are symmetric with respect to those
given in Eqs. 22 and 23. At present,  this is the $n=1$ LL
attached to the valley K (in our convention) which starts to be
filled and the notation $f_{1}^{\pm}$ refer to this LL. Of course
some of the diagonal matrix elements are changed  but results are
formally similar and the corresponding exciton dispersion curves
are displayed in Fig.6.  All the eigen-value solutions of Fig.6
can be expressed analytically:

For $4<\nu<5$ one gets one solution $Ed2(K)= h_{55}(K)$ which is
not optically active and two groups of optically active solutions:
\begin{equation}
\begin{split}
E0_{4<\nu<5}^{\pm}&=\frac{1}{2}[h_{11}+h_{22}\\
&\pm\sqrt{(h_{11}-h_{22})^{2}+ 4 V_{0101}^{2}(1-f_{1}^{-})}]
\end{split}
\end{equation}
and:

\begin{equation}
\begin{split}
E1_{4<\nu<5}^{\pm}&=\frac{1}{2}[h_{33}+h_{44}+ V_{1212}\\
&\pm\sqrt{(h_{44}-h_{33}+ V_{1212})^{2}+ 8 V_{1212}^{2}f_{1}^{-}}]
\end{split}
\end{equation}
for which all matrix elements, dependent on $K$, are given in
appendix B (Eq. B11).

For $5<\nu<6$ one gets two solutions $Ed2(K)= h_{44}(K)$ (same
expression as for $4<\nu<5$) which are not optically active, one
optically active solution $E0_{5<\nu<6}(K)= h_{11}(K) $ and two
other optically active solutions denoted as $E1_{5<\nu<6}^{\pm}$:

\begin{equation}
\begin{split}
E1_{5<\nu<6}^{\pm}&=\frac{1}{2}[h_{22}+h_{33}+2V_{1212}\\
&\pm\sqrt{(h_{22}-h_{33}-2V_{1212})^{2}+12 V_{1212}^{2}f_{1}^{+}}]
\end{split}
\end{equation}
The related matrix elements, dependent on $K$, are given in
appendix B (Eq. B12).

We will not discuss the case of interband transitions for this
configuration of filling factors but their corresponding
Hamiltonians $\widetilde{HI^{12}_{2<\nu<6}}$ and
$\widetilde{HI^{23}_{2<\nu<6}}$ are modified in two respects: for
both of them the exchange contributions entering the diagonal
elements are different and for $\widetilde{HI^{12}_{2<\nu<6}}$ the
transitions are now filling factor dependent in such a way the
corresponding transitions disappear at $\nu =6$.

The case of filling factors $6<\nu <10$ and following ones will
not be discussed as well but the corresponding treatment is
formally similar to the case $2<\nu <6$ with a different set of
one electron energies, exchange contributions and matrix elements.

We are now focussing the discussion on the results obtained for
$K\simeq 0$ which could be compared to magneto-optical absorption
measurements.

\section{ Discussion of the results for $ K \simeq 0 $}

For $K=|\overrightarrow{k}l_{B}|\simeq 0$, all the Hamiltonians
are reduced to their diagonal elements which are given in appendix
B. The reason is that all off-diagonal elements are  proportional
to $K$ or $K^{2}$. We will call the corresponding solutions, at
$K\simeq 0$, $E_{MP}^{n,n+1}$ and $EI_{MP}^{|n|,|n+1|}$ for
intra-LL transitions and interband transitions respectively. We
restrict the discussion to those solutions which are optically
active. All results are function of
$\alpha_{0}=\frac{1}{2}\sqrt{\frac{\pi}{2}}= 0.627$ in Coulomb
units. We then get the following results:

For the transitions $E_{MP}^{01}$ which involve the $n=0$ LL:

\begin{align}
E_{MP}^{01} &= E_{10} + C_{1}- \frac{3}{4}\alpha_{0}\nonumber\\
&(\text{for}\hspace{0.1cm} \nu=1,2,3,4,5) \nonumber\\
E_{MP}^{01} &= E_{10}+ C_{1} +
\frac{\alpha_{0}}{4}(-3+\frac{5}{2}(6-\nu)) \nonumber\\
&(\text{for}\hspace{0.1cm}  5<\nu<6)
\end{align}

As clearly apparent in Fig. 6 and 7, this transition is split for
non integer values of $\nu> 2$. The  high energy component of this
split level has an energy which increases with $\nu$ but its
oscillator strength decreases with $\nu$ going to zero at integer
value of $\nu$.

For the transitions $E_{MP}^{12}$ which involve the transitions
between the $n=1$ and $n=2$ LL:

\begin{align}
E_{MP}^{12} &= (\sqrt2-1)(E_{10}+C_{1})+ \Delta C_{2}\nonumber\\
&-\frac{\alpha_{0}}{16}(1+ 2\sqrt2)\nonumber\\
&(\text{for}\hspace{0.1cm} \nu=3,4,5,6) \nonumber\\
E_{MP}^{12} &= (\sqrt2-1)(E_{10}+C_{1})+ \Delta C_{2}\nonumber\\
&+\frac{\alpha_{0}}{16}(-14+(13- 2\sqrt2)(\nu-2))\nonumber\\
&(\text{for}\hspace{0.1cm} 2<\nu<3)
\end{align}

 In this case also this transition is split for non
integer values of $\nu$. The  high energy component of this split
level has an energy which increases with $\nu$ but its oscillator
strength decreases with $\nu$ going to zero at integer value of
$\nu$.

The optical active interband transitions $EI_{MP}^{12}$ which
involve the transitions between the $n=-2,-1$ and $n=1,2$ Landau
levels are split by an amount $\frac{\alpha_{0}}{8}$ but the mean
energy $EIm_{MP}^{12}=(EI_{1}^{+}(0)+EI_{1}^{-}(0))/2$ (Eq. 19)
has the following expression:

\begin{equation}
EIm_{MP}^{12}= (\sqrt{2}+1)(E_{10}+ C_{1})+ \Delta
C_{2}-\frac{33}{32}\alpha_{0}
\end{equation}
 whereas the  mean energy for the optically active transition
implying the $n=-3,-2$ and $n=2,3$ Landau levels is expressed as:

\begin{equation}
EIm_{MP}^{23}= (\sqrt{3}+\sqrt{2})(E_{10}+ C_{1})+ \Delta
C_{3}-\frac{233}{256}\alpha_{0}
\end{equation}

For a given value of the Fermi velocity $v_{F}$ ( here equal to
$0.86\times 10^{6}m/s$) the energy $E_{10}$ is determined (here
$E_{10}= 2.77$) and in Eqs. 30 to 33 the only unknown parameter
is $C_{1}$.

As already said this quantity defined in appendix A (Eq. A2) is
divergent. The occurrence of such a problem is not specific of the
Graphene properties because it is also present in C2DEG though it
was not explicitly formulated. The reason why this term appears
here is that we wanted to define a \textbf{common }energy scale
for intraband and interband transitions. This will corresponds in
C2DEG to impose a common energy scale to cyclotron-type
transitions and inter-band excitonic transitions. There was an
attempt to treat this later transitions in GaAs, in an another
context, but using the same theoretical model \cite{By4} and
indeed the same problem of divergence of the exchange interaction
among the valence band levels was found without being able to
solve it. Therefore this problem is not specific to Graphene but,
in that case, one can solve it, at least, in a semi-empirical way.

The divergence of $C_{1}$ is due to the infinite summation over LL
(see Eq. A2) which is physically artificial. We could then define,
 as was done in Ref. \cite{Iyengar}, a cut-off value on energy or number of LL, but this limit is quite
arbitrary. We propose to treat the problem in a semi-empirical
way, using $C_{1}$ as a parameter fitted, for one type of
transitions, to experimental data and then to deduce all the
re-normalized velocities attached to the other transitions. Doing
so, we \textit{implicitly assume }that all the re-normalization of
the velocity, for the fitted transition,  is only due  to
electron-electron interactions neglecting any possible
contribution from electron-phonon interaction which may be
important in carbon based compounds. Among experimental data which
could be used for this fitting, those related to
magneto-transmission measurements  \cite{Sadowski,Jiang} are those
which are expected to reflect the magneto-plasmon picture
developed is this study. Another set of data, based on
photoconductivity measurements  on ex-foliated Graphene
\cite{Deacon} can also be considered to compare results. We will
use the data of Ref.\cite{Jiang}, obtained on ex-foliated
Graphene, to fit $C_{1}$ to the $E_{MP}^{01}$ transition at
$\nu=2$. In this reference, the re-normalized Fermi velocity
$\widetilde{c}_{01}^{ex}= (1.12\pm 0.02)\times 10^{6}m/sec$ which,
from Eq.30, leads to a value $C_{1}= 1.31 \pm 0.06$. When
injecting this value in Eq. 31, we predict a re-normalized Fermi
velocity $\widetilde{c}I_{12}^{th}= (1.163\pm 0.02)\times
10^{6}m/sec$ to be compared with the corresponding experimental
value \cite{Jiang} $\widetilde{c}I_{12}^{ex}= (1.18\pm 0.02)\times
10^{6}m/sec$. The agreement is reasonable. As seen on both
experimental and theoretical grounds, the re-normalized  velocity
differs for different transitions. One can then try to evaluate
these velocities for other transitions. The results are given in
Table 1 for integer values of $\nu$. The value quoted in the
table, from Ref.\cite{Deacon}, corresponds to the transition $n=0$
to $n=1$ whereas the corresponding value for the transition $n=-1$
to $n=0$ is found to be $(1.07 \pm 0.004)\times 10^{6}m/sec$. This
corresponds to an asymmetry of the conduction and valence levels
not taken into account in our model but also not reported in
Ref.\cite{Jiang} for this transition.

In table 1, we have included the results of the re-normalized
Fermi velocity for transitions between LL $n=3$ to $n=4$, $n=2$ to
$n=3$, $n=-3$ to $n=4$ for which we have calculated the diagonal
elements of the corresponding Hamiltonians.

\begin{table}
\caption{\label{Table 1} Evaluation of the re-normalized
velocities $\widetilde{c}$, at integer values of the filling
factor, for different transitions. }
\begin{ruledtabular}
\begin{tabular}{cccc}
Transition & $\nu$ & $\widetilde{c}^{ex}(10^{6}m/s)$ & $\widetilde{c}^{th}(10^{6}m/s)$\\
$n$ to $m$ & &   &  \\
\hline
 $3$ to $4$ & $8,10$ &           &$0.99 \pm 0.02$ \\
 $2$ to $3$ & $6,8$  &           &$1.01 \pm 0.02$ \\
 $1$ to $2$ & $4,6$  &           &$1.04 \pm 0.02$ \\
 $0$ to $1$ & $2,4$  &$1.12 \pm 0.02 $ \footnotemark [1]&$1.12 \pm 0.02$ \\
            & $2$    &$1.12 \pm 0.004$ \footnotemark [2]&\\
 $-1$ to $2$& $2$    &$1.18 \pm 0.02$ \footnotemark [1]&$1.16 \pm 0.02$ \\
$-2$ to $3$ &$2$     &          &$ 1.16 \pm 0.02 $\\
 $-3$ to $4$& $2$    &          &$1.16 \pm 0.02$ \\
\end{tabular}
\end{ruledtabular}
\footnotetext [1]{from Ref. \cite{Jiang}} \footnotetext [2]{from
Ref. \cite{Deacon}}
\end{table}

As clearly apparent, from Table 1, $\widetilde{c}$ varies strongly
with the transition though it seems to be relatively constant for
all interband transitions. This qualitative feature is also
observed in experiments performed on epitaxial multi-layer
Graphene \cite{Sadowski,Sadowski1}. However in these experiments
the reported values of $\widetilde{c}^{ex}$ for all transitions is
the same and equal to $ (1.03 \pm 0.01)\times 10^{6}m/s$ which is,
at present, not understood.

One could, a priori, think that the results obtained are dependent
on the value of $v_{F}$ adopted in the calculations. In fact one
can vary this value over a large range, for instance, from $0.80
\times 10^{6}m/s$ to $0.90 \times 10^{6}m/s$ getting values for
$C_{1}= 1.50$ to $1.18$ respectively but the quantity which enters
the Hamiltonians  is in fact $E_{01}+ C_{1}$ which \textbf{remains
constant}, independent on $v_{F}$ and equal to $4.08$. This value
has been adopted to calculate the dispersion curves of Figs. 2, 3,
4, 6 and 7. It is therefore not possible from experimental data on
energies to determine $C_{1}$ but all transitions are now given
with a \textbf{common} energy scale. On the other hand the
oscillator strengths of the transitions are proportional to
$v_{F}^{2}$ and then absolute transmission measurements could in
principle give information on $v_{F}$ and therefore on $C_{1}$.

\section{ Conclusions}

In conclusion, we have developed  , within the Hartree-Fock
approximation, a full treatment of the magnetoplasmon picture in
Graphene valid for a very large range of  magnetic fields. This
model, applied for filling factors up to 6, shows that the
electron-electron interactions induce different effects: (i) for
some of the transitions these interactions lead to a splitting of
the optical transitions and (ii) they are responsible for a strong
re-normalization of the Fermi velocity as observed in
magneto-optical experiments. This re-normalization is found to be
dependent  on the type of investigated transitions. The optical
conductivity components have been evaluated showing that the
oscillator strength of the optical transitions is proportional to
$v_{F}^{2}$ and not to
  the square of the re-normalized velocity.The theory has been derived
  for all transitions with a common
energy scale which should allow a direct comparison of its
predictions with future experimental works.

\section{Acknowledgments}

The GHMFL is "Laboratoire conventionn\'{e}  \`{a }l'UJF et l'INPG
de Grenoble". The work presented here has been supported in part
by the European Commission through the Grant RITA-CT-2003-505474.

\appendix
\section{ Exchange contributions}

 We report in this appendix the explicit expressions
for the contribution of the exchange energies entering the
diagonal elements of the different Hamiltonian matrices in units
of Coulomb energies. We introduce the notation
$\alpha_{0}=\frac{1}{2}\sqrt\frac{\pi}{2}$ which characterize the
exchange interaction in C2DEG at $\nu=1$. Applying the expression
given in Eq. 12 we obtain successively the contribution of
exchange for the different Hamiltonians. To simplify the notations
we will drop the superscript
$\hspace{0.1cm}\widetilde{}\hspace{0.1cm}$   from
$\widetilde{E}_{n,m,n,m}(0)$ meaning that all these quantities are
real.

\subsection{ Exchange contributions to
$\widetilde{H_{1<\nu<2}^{0}}$ and $\widetilde{H_{0<\nu<1}^{0}}$ }

For $\widetilde{H_{1<\nu<2}^{0}}$ we obtain:
\begin{align}
\sum_{m}(E_{0,m,0,m}(0)-E_{1,m,1,m}(0))f_{m}&=\nonumber\\
\frac{3}{4}\alpha_{0}(2f_{0}^{+}-1)+ C_{1}\nonumber\\
\sum_{m}(E_{0,m,0,m}(0)-E_{1,m,1,m}(0))f_{m}^{-}&=
\frac{3}{4}\alpha_{0}+ C_{1}\nonumber\\
\sum_{m}(E_{-1,m,-1,m}(0)-E_{0,m,0,m}(0))f_{m}^{+}&=\nonumber\\
-\frac{3}{4}\alpha_{0}(2f_{0}^{+}-1)+ C_{1}
\end{align}
 where:
\begin{equation}
C_{1}=  \frac{1}{\sqrt2}\sum_{m}\int_{0}^{\infty}dx
e^{-x^{2}}\frac{x^{2}}{\sqrt{m+1}}L_{m}^{1}
\end{equation}

and $L_{m}^{\alpha}$ are Laguerre polynomials of argument $x^{2}$
in this Appendix.

The quantity $C_{1}$ diverges due to the simplifying assumption of
the
 infinite linear dispersion of the Graphene bands. The summation
 has to be truncated at some level or this parameter has to be
 fitted to experimental data (see section 5).

For $\widetilde{H_{0<\nu<1}^{0}}$ we obtain for the exchange part
the same expressions than those  given in Eq.A1  replacing
$f_{0}^{+}$ by $f_{0}^{-}$.

\subsection{  Exchange contributions to $\widetilde{H_{1<\nu<2}^{12}}$ and $\widetilde{H_{0<\nu<1}^{12}}$}

For $\widetilde{HI_{1<\nu<2}^{12}}$ one gets:

\begin{align}
\sum_{m}(E_{-1,m,-1,m}(0)-E_{2,m,2,m}(0))f_{m}^{-}&=
\frac{\alpha_{0}}{16}+ CI_{12}\nonumber\\
\sum_{m}(E_{-2,m,-2,m}(0)-E_{1,m,1,m}(0))f_{m}^{-}&=
-\frac{\alpha_{0}}{16}+ CI_{12}\nonumber\\
\sum_{m}(E_{-1,m,-1,m}(0)-E_{2,m,2,m}(0))f_{m}^{+}&=\nonumber\\
\frac{\alpha_{0}}{16}(2f_{0}^{+}-1)+ CI_{12}\nonumber\\
\sum_{m}(E_{-2,m,-2,m}(0)-E_{1,m,1,m}(0))f_{m}^{+}&=\nonumber\\
-\frac{\alpha_{0}}{16}(2f_{0}^{+}-1)+ CI_{12}
\end{align}
where:
\begin{align}
CI_{12}&= \frac{1}{\sqrt{2}}\sum_{m=0}^{\infty}\int_{0}^{\infty}dx
e^{-x^{2}}\frac{x^{2}}{\sqrt{m+1}}L_{m}^{1}[1+\frac{L_{1}^{1}}{\sqrt{2}}]\nonumber\\
&=(\sqrt2+1)C_{1}+ \Delta C_{2}\nonumber\\
\Delta C_{2}&= -\frac{1}{2}\sum_{m=0}^{\infty}\int_{0}^{\infty}dx
e^{-x^{2}}\frac{x^{4}}{\sqrt{m+1}}
\end{align}
 $CI_{12}$ in this equation also diverges  like
$C_{1}$ but  $\Delta C_{2}$ converges to a value -0.156.

Similar expressions hold for $\widetilde{HI_{0<\nu<1}^{12}}$ when
replacing $f_{0}^{+}$ by $f_{0}^{-}$.

\subsection{  Exchange contributions to  $\widetilde{H_{1<\nu<2}^{23}}$ and $\widetilde{H_{0<\nu<1}^{23}}$}

For $\widetilde{HI_{1<\nu<2}^{23}}$ one gets:

\begin{align}
\sum_{m}(E_{-2,m,-2,m}(0)-E_{3,m,3,m}(0))f_{m}^{-}&=
\frac{\alpha_{0}}{32}+ CI_{23}\nonumber\\
\sum_{m}(E_{-3,m,-3,m}(0)-E_{2,m,2,m}(0))f_{m}^{-}&=
-\frac{\alpha_{0}}{32}+ CI_{23}\nonumber\\
\sum_{m}(E_{-2,m,-2,m}(0)-E_{3,m,3,m}(0))f_{m}^{+}&=\nonumber\\
\frac{\alpha_{0}}{32}(2f_{0}^{+}-1)+ CI_{23}\nonumber\\
\sum_{m}(E_{-3,m,-3,m}(0)-E_{2,m,2,m}(0))f_{m}^{+}&=\nonumber\\
-\frac{\alpha_{0}}{32}(2f_{0}^{+}-1)+ CI_{23}
\end{align}
where:
\begin{align}
CI_{23}&=  \frac{1}{\sqrt2}\sum_{m}\int_{0}^{\infty}dx
e^{-x^{2}}\frac{x^{2}}
{\sqrt{m+1}}L_{m}^{1}(\frac{L_{1}^{1}}{\sqrt2}+\frac{L_{2}^{1}}{\sqrt3})\nonumber\\
&=(\sqrt{3}+\sqrt{2})C_{1}+ \Delta C_{3}\nonumber\\
\Delta C_{3}&= -\frac{1}{2}\sum_{m=0}^{\infty}\int_{0}^{\infty}dx
e^{-x^{2}}\frac{x^{4}}{\sqrt{m+1}}(1+\sqrt6-\frac{x^{2}}{\sqrt6})
\end{align}
 $CI_{23}$ in this equation also diverges  like
$C_{1}$ but  $\Delta C_{3}$ converges to a value -0.467.

Similar expressions hold for $\widetilde{HI_{0<\nu<1}^{23}}$ when
replacing $f_{0}^{+}$ by $f_{0}^{-}$.

Comparing Eqs. A4 and A6 one can formally extend the treatment and
find that, for any interband transition from LL $-p$ to LL $q =
p+1$ the corresponding  divergent term $CI_{pq}$ entering the
exchange contributions is given by $CI_{pq}=
(\sqrt{p}+\sqrt{q})C_{1})+ F_{p,q}$ where $F_{p,q}$ is finite.

\subsection{  Exchange contributions to
$\widetilde{H_{2<\nu<6}^{12}}$}

The different contributions to the exchange for the different
Hamiltonians are:

\begin{align}
\sum_{m}(E_{0,m,0,m}(0)-E_{1,m,1,m}(0))f_{m}^{\pm}&=\nonumber\\
\alpha_{0}(\frac{3}{4}-\frac{7}{8}f_{1}^{\pm})+ C_{1}\nonumber\\
\sum_{m}(E_{1,m,1,m}(0)-E_{2,m,2,m}(0))f_{m}^{\pm}&=\nonumber\\
\frac{\alpha_{0}}{32}(5+ (26- 4 \sqrt{2})f_{1}^{\pm})+ C_{12}
\end{align}
where $C_{12}=(\sqrt2-1)C_{1}+ \Delta C_{2}$.

Here also  formally, when extending the treatment, one finds that,
for any intra-LL transition from LL $p$ to LL $q = p+1$ the
corresponding divergent term $C_{pq}$ entering the exchange
contributions is given by $C_{pq}= (\sqrt{q}-\sqrt{q})C_{1})+
G_{p,q}$ where $G_{p,q}$ is finite. Note however that in general
$G_{p,q}$ is different from $F_{p,q}$.

\section{  Hamiltonian matrix elements}

We report in this appendix the explicit expressions for the matrix
elements of the Hamiltonian matrices in units of Coulomb energies.
Results are given as a function of $K= |\overrightarrow{k}l_{B}|$.
For simplicity we adopt the same notation $h_{ij}$ for noting the
matrix elements of all matrices but their expression is specific
of the case under consideration. All matrix elements
$\widetilde{\widetilde{V}}_{n_{1},n_{2},n_{3},n_{4}}(\overrightarrow{q})$
and $\widetilde{E}_{n_{1},n_{2},n_{3},n_{4}}(\overrightarrow{k})$
are evaluated using Eq. 11.

\subsection{ Matrix elements of $\widetilde{H_{1<\nu<2}^{0}}$ and $\widetilde{H_{0<\nu<1}^{0}}$ }

 For $\widetilde{H_{1<\nu<2}^{0}}$ we obtain:

\begin{equation}
\begin{split}
h_{11}&= E_{10}+\frac{3}{4}\alpha_{0}(2f_{0}^{+}-1)+
f_{0}^{+}(V_{0101}-E_{0110})+ C_{1}\\
h_{22}&=h_{33}=h_{44}=E_{10}+\frac{3}{4}\alpha_{0}+
(V_{0101}-E_{0110})+ C_{1}\\
h_{55}&=E_{10}-\frac{3}{4}\alpha_{0}(2f_{0}^{+}-1)
+(1-f_{0}^{+})\\
 &\times(V_{-10-10}-E_{-100-1})+ C_{1}
\end{split}
\end{equation}


The matrix elements $V_{n_{1}n_{2}n_{3}n_{4}}$ entering the
Hamiltonian $\widetilde{H_{1<\nu<2}^{0}}$ are:
\begin{equation}
V_{0101}(K)= V_{0011}(K)=\frac{K}{4}e^{-\frac{K^{2}}{2}}
\end{equation}
with $\widetilde{\widetilde{V}}_{0,1,0,1}= V_{0101}$
 and
$\widetilde{\widetilde{V}}_{0,0,1,1}= V_{0011}e^{2\imath\varphi}$
where $\varphi$ is the polar angle of the exciton wave vector.

The matrix elements $E_{n_{1}n_{2}n_{3}n_{4}}$ entering Eqs. 15
and B1 are:

\begin{equation}
\begin{split}
E_{0110}(K)&=\\
&\sqrt{\frac{\pi}{2}}[\Phi(\frac{1}{2},1;-\frac{K^{2}}{2})-\frac{1}{4}\Phi(\frac{3}{2},1;-\frac{K^{2}}{2})] \\
E_{0011}(K)&=-\frac{3K^{2}}{32}\sqrt{\frac{\pi}{2}}\Phi(\frac{5}{2},3;-\frac{K^{2}}{2})
\end{split}
\end{equation}

where $\widetilde{E}_{0,1,1,0}=E_{0110}$,
$\widetilde{E}_{0,0,1,1}=E_{0011}e^{2\imath\varphi}$
 and
$\Phi(a,b;z)$ is the confluent hypergeometric function.

For $\widetilde{H_{0<\nu<1}^{0}}$ the matrix elements are
identical to those given in Eqs. B1, B2 and B3 replacing
$f_{0}^{+}$ by $f_{0}^{-}$ when appropriate.

\subsection{ Matrix elements of $\widetilde{HI_{1<\nu<2}^{12}}$ and
$\widetilde{HI_{0<\nu<1}^{12}}$ }

The matrix elements $h_{ij}$ of $\widetilde{HI_{1<\nu<2}^{12}}$
are:
\begin{align}
h_{11}&= EI_{12}+\frac{\alpha_{0}}{16}+ C_{2}+V_{-12-12}-E_{-122-1}\notag\\
h_{22}&= EI_{12}-\frac{\alpha_{0}}{16}+ C_{2}+V_{-12-12}-E_{-122-1}\notag\\
h_{33}&= EI_{12}+\frac{\alpha_{0}}{16}(2f_{0}^{+}-1)+ C_{2}+V_{-12-12}-E_{-122-1}\notag\\
h_{44}&= EI_{12}-\frac{\alpha_{0}}{16}(2f_{0}^{+}-1)+ C_{2}+V_{-12-12}-E_{-122-1}\notag\\
h_{55}&= EI_{12}+\frac{\alpha_{0}}{16}+ C_{2}+V_{-12-12}-E_{-122-1}\notag\\
h_{66}&= EI_{12}-\frac{\alpha_{0}}{16}+ C_{2}+V_{-12-12}-E_{-122-1}\notag\\
h_{77}&= EI_{12}+\frac{\alpha_{0}}{16}+ C_{2}+V_{-12-12}-E_{-122-1}\notag\\
h_{88}&= EI_{12}-\frac{\alpha_{0}}{16}+ C_{2}+V_{-12-12}-E_{-122-1}\notag\\
h_{12}&=h_{34}=h_{56}=h_{78}=V_{-11-22}-E_{-112-2}\notag\\
h_{13}&=h_{15}=h_{17}=h_{24}=h_{26}=h_{28}= V_{-12-12}\notag\\
h_{35}&=h_{37}=h_{46}=h_{48}=h_{58}=h_{68}= V_{-12-12}\notag\\
h_{14}&=h_{16}=h_{18}=h_{23}=h_{25}=h_{27}= V_{-11-22}\notag\\
h_{36}&=h_{38}=h_{45}=h_{47}=h_{57}=h_{67}=V_{-11-22}
\end{align}
where $EI_{12}=(\sqrt2+1)E_{10}$. The matrix elements
$V_{n_{1}n_{2}n_{3}n_{4}}$ entering Eq. B4 are:

\begin{equation}
\begin{split}
V_{-12-12}(K)&=V_{-11-22}(K)=\frac{K}{8}e^{-\frac{K^{2}}{2}}
[(3-2\sqrt{2})\\
&+(\sqrt{2}-2)\frac{K^{2}}{2}+\frac{K^{4}}{8}]\\
\end{split}
\end{equation}

with $\widetilde{\widetilde{V}}_{-1,2,-1,2}= V_{-12-12}$
 and
$\widetilde{\widetilde{V}}_{-1,1,-2,2}=
V_{-11-22}e^{2\imath\varphi}$.

The matrix elements $E_{n_{1}n_{2}n_{3}n_{4}}$ entering B4 are:

\begin{multline}
E_{-122-1}(K)=\sqrt{\frac{\pi}{2}}[\Phi(\frac{1}{2};1;-\frac{K^{2}}{2})\\
-\Phi(\frac{3}{2};1;-\frac{K^{2}}{2})+\frac{3}{4}\Phi(\frac{5}{2};1;-\frac{K^{2}}{2})-\frac{15}{64}\Phi(\frac{7}{2};1;-\frac{K^{2}}{2})]\\
E_{-112-2}(K)=
-\frac{K^{2}\sqrt{\pi}}{64}[\frac{3(1+\sqrt{2})^{2}}{\sqrt{2}}
\Phi(\frac{5}{2};3;-\frac{K^{2}}{2})\\
-15(2+\sqrt{2})\Phi(\frac{7}{2};3;-\frac{K^{2}}{2})
+\frac{105\sqrt{2}}{16}\Phi(\frac{9}{2};3;-\frac{K^{2}}{2})]
\end{multline}

where $\widetilde{E}_{-1,2,2,-1}=E_{-122-1}$ and
$\widetilde{E}_{-1,1,2,-2}=E_{-112-2}e^{2\imath\varphi}$.

For $\widetilde{HI_{0<\nu<1}^{12}}$ two columns of  the matrix
$\widetilde{HI_{1<\nu<2}^{12}}$ are inverted but the eigen values
are the same with $f_{0}^{-}$ replacing $f_{0}^{+}$ in Eq. B6.

\subsection{ Matrix elements of $\widetilde{HI_{1<\nu<2}^{23}}$ and
$\widetilde{HI_{0<\nu<1}^{23}}$ }

The matrix elements $h_{ij}$ of $\widetilde{HI_{1<\nu<2}^{23}}$
are similar to those given in Eq. B4 when replacing $V_{-12-12}$,
$V_{-11-22}$, $E_{-122-1}$, $E_{-112-2}$ by $V_{-23-23}$,
$V_{-22-33}$, $E_{-233-2}$, $E_{-223-3}$ respectively and
$EI_{12}$ by $EI_{23}= (\sqrt3+\sqrt2)E_{10}$. The new
 matrix elements
$E_{n_{1}n_{2}n_{3}n_{4}}$ are here:
\begin{equation}
\begin{split}
E_{-233-2}(K)&=\sqrt{\frac{\pi}{2}}[\Phi(\frac{1}{2};1;-\frac{K^{2}}{2})-2\Phi(\frac{3}{2};1;-\frac{K^{2}}{2})\\
&+\frac{15}{4}\Phi(\frac{5}{2};1;-\frac{K^{2}}{2})-\frac{795}{192}\Phi(\frac{7}{2};1;-\frac{K^{2}}{2})\\
&+
\frac{315}{128}\Phi(\frac{9}{2};1;-\frac{K^{2}}{2})-\frac{945}{1536}\Phi(\frac{11}{2};1;-\frac{K^{2}}{2})]\\
 E_{-223-3}(K)&=-\frac{K^{2}\sqrt{\pi}}{64}[3(5+2\sqrt6)
\Phi(\frac{5}{2};3;-\frac{K^{2}}{2})\\
&-15(4+\frac{3\sqrt3}{\sqrt{2}})\Phi(\frac{7}{2};3;-\frac{K^{2}}{2})\\
&+105(\frac{9}{8}+\frac{\sqrt{6}}{3})\Phi(\frac{9}{2};3;-\frac{K^{2}}{2})\\
&-945\frac{6+\sqrt6}{48}\Phi(\frac{11}{2};3;-\frac{K^{2}}{2})\\
&+\frac{10395}{192}\Phi(\frac{13}{2};3;-\frac{K^{2}}{2})]
\end{split}
\end{equation}

where $\widetilde{E}_{-2,3,3,-2}=E_{-233-2}$ and
$\widetilde{E}_{-2,2,3,-3}=E_{-223-3}e^{2\imath\varphi}$.

 and the corresponding matrix elements $V_{n_{1}n_{2}n_{3}n_{4}}$:

\begin{equation}
\begin{split}
V_{-23-23}(K)&=V_{-22-33}(K)=\frac{K}{8}e^{-\frac{K^{2}}{2}}
[(3-2\sqrt{6})\\
& +(\frac{3\sqrt6}{2}-8)K^{2}
+(\frac{9}{8}-\frac{\sqrt6}{3})K^{4}\\
&+\frac{(\sqrt6-6)}{48}K^{6}+\frac{K^{8}}{192}]
\end{split}
\end{equation}

with $\widetilde{\widetilde{V}}_{-2,3,-2,3}= V_{-23-23}$
 and
$\widetilde{\widetilde{V}}_{-2,2,-3,3}=
V_{-22-33}e^{2\imath\varphi}$.

\subsection{ Matrix elements of $\widetilde{H_{2<\nu<6}^{12}}$}

 For $\widetilde{H_{2<\nu<3}^{12}}$ we obtain:

\begin{equation}
\begin{split}
h_{11}&=h_{22}=h_{33}=E_{10}+\frac{3}{4}\alpha_{0}+
(V_{0101}-E_{0110})+ C_{1}\\
h_{44}&= E_{10}+ \alpha_{0}(\frac{3}{4}-\frac{7}{8}f_{1}^{-})+
(1-f_{1}^{-})(V_{0101}-E_{0110})+ C_{1}\\
h_{55}&=E_{12}+\frac{\alpha_{0}}{16}(1+\frac{57-30\sqrt2}{6})f_{1}^{-}\\
&+f_{1}^{-} (V_{1212}-E_{1221})+ C_{2}^{'}
\end{split}
\end{equation}

For $\widetilde{H_{3<\nu<4}^{12}}$ :

\begin{equation}
\begin{split}
h_{11}&=h_{22}=E_{10}+\frac{3}{4}\alpha_{0}+
(V_{0101}-E_{0110})+ C_{1}\\
h_{33}&= E_{10}+ \alpha_{0}(\frac{3}{4}-\frac{7}{8}f_{1}^{+})\\
&+(1-f_{1}^{+})(V_{0101}-E_{0110})+ C_{1}\\
h_{44}&=E_{12}+\frac{\alpha_{0}}{32}(21-10\sqrt2)\\
&+(V_{1212}-E_{1221})+ C_{2}^{'}\\
h_{55}&=E_{12}+\frac{\alpha_{0}}{16}(1+\frac{19-10\sqrt2}{2})f_{1}^{+}\\
&+f_{1}^{+} (V_{1212}-E_{1221})+ C_{2}^{'}
\end{split}
\end{equation}

For $\widetilde{H_{4<\nu<5}^{12}}$ :

\begin{equation}
\begin{split}
h_{11}&=E_{10}+\frac{3}{4}\alpha_{0}+
(V_{0101}-E_{0110})+ C_{1}\\
h_{22}&= E_{10}+ \alpha_{0}(\frac{3}{4}-\frac{7}{8}f_{1}^{-})\\
&+(1-f_{1}^{-})(V_{0101}-E_{0110})+ C_{1}\\
h_{33}&=E_{12}+\frac{\alpha_{0}}{16}(1+\frac{19-10\sqrt2}{2})f_{1}^{-}\\
&+f_{1}^{-} (V_{1212}-E_{1221})+ C_{2}^{'}\\
h_{44}&= h_{55}=E_{12}+\frac{\alpha_{0}}{32}(21-10\sqrt2)\\
&+(V_{1212}-E_{1221})+ C_{2}^{'}
\end{split}
\end{equation}

For $\widetilde{H_{5<\nu<6}^{12}}$ :

\begin{equation}
\begin{split}
h_{11}&= E_{10}+ \alpha_{0}(\frac{3}{4}-\frac{7}{8}f_{1}^{+})\\
&+(1-f_{1}^{-+})(V_{0101}-E_{0110})+ C_{1}\\
h_{22}&=E_{12}+\frac{\alpha_{0}}{16}(1+\frac{19-10\sqrt2}{2})f_{1}^{+}\\
&+f_{1}^{+} (V_{1212}-E_{1221})+ C_{2}^{'}\\
h_{33}&= h_{44}= h_{55}= E_{12}+\frac{\alpha_{0}}{32}(21-10\sqrt2)\\
&+(V_{1212}-E_{1221})+ C_{2}^{'}
\end{split}
\end{equation}
with the corresponding new matrix elements entering Eq. B9, B10,
B11 and B12:
\begin{align}
V_{1212}(K)&=\frac{K}{8}e^{-\frac{K^{2}}{2}}[1+\sqrt2-\frac{K^{2}}{2\sqrt2}]^{2}\nonumber\\
E_{1221}(K)&=E_{-122-1}(K)
\end{align}
where $\widetilde{\widetilde{V}}_{1,2,1,2}= V_{1212}$

\section{  Optical conductivity}
In this appendix we derive, following the lines of Ref.\cite{By2},
the corresponding expressions which allow to calculate the optical
matrix elements of the MP curves which enter in the optical
conductivity $\overline{\overline{\sigma}}(\hbar\omega)$ which has
two components:
\begin{align}
\sigma_{\parallel}&=-\imath
\frac{e^{2}G_{B}}{\omega}\sum_{j}\frac{2E_{MP}^{j}|
\overrightarrow{{\cal{M}}_{\parallel}}\cdot\overrightarrow{L_{j}}|^{2}}
{(E_{MP}^{j})^{2}-(\hbar\omega)^{2}}\nonumber\\
\sigma_{\perp} &= -\imath\frac{e^{2}G_{B}}{\omega}
\sum_{j}\frac{2\hbar\omega(\overrightarrow{{\cal{M}}_{\parallel}}\cdot\overrightarrow{L_{j}})
(\overrightarrow{{\cal{M}}_{\perp}}\cdot\overrightarrow{L_{j}})^{*}}{(E_{MP}^{j})^{2}-(\hbar\omega)^{2}}
\end{align}
where the summation is performed on all MP transitions of energy
$E_{MP}^{j}$ with the corresponding eigen vector
$\overrightarrow{L_{j}}$. In Eq. C1, $G_{B}=1/(2\pi (l_{B})^{2})$
is the density of states of a single LL,
$\overrightarrow{{\cal{M}}_{\parallel}}$ and
$\overrightarrow{{\cal{M}}_{\perp}}$ are optical vectors with
components
 $M_{aj}^{-1}{\cal{F}}_{\parallel;aj}^{i}$ and
 $M_{aj}^{-1}{\cal{F}}_{\perp;aj}^{i}$ respectively.
 $\widehat{M}$ is the matrix used to symmetrize the Hamiltonian
 and depends on the set of transitions which are considered (see Eqs. 16, 18, 21).
$aj$ denotes one of the transition $n$ to $m$ belonging to this
set of transitions. ${\cal{F}}_{\alpha ;aj}$  with $\alpha= \perp,
\parallel $ is defined in reduced units
($\overrightarrow{K}$ standing for $\overrightarrow{k}l_{B}$ and
$u$ for $x/l_{B}$) as:

\begin{equation}
{\cal{F}}_{\alpha; m,n}(\overrightarrow{K})=\int du e^{\imath
K_{x}u}
[(F_{m}^{i}(u+\frac{K_{y}}{2}))^{*}\widehat{V_{\alpha}^{i}}F_{n}^{i}(u-\frac{K_{y}}{2}))]
\end{equation}
where the function $[F^{*}\widehat{V}F]$ denotes the scalar
product and the velocity operators $\widehat{V_{\alpha}^{i}}$ are:
\begin{align}
\widehat{V_{\parallel}^{K}}&= v_{F}
\begin{bmatrix}0&-\imath e^{-\imath\varphi}\\
\imath e^{\imath\varphi}&0\end{bmatrix}\nonumber\\
\widehat{V_{\perp}^{K}}&= v_{F}
\begin{bmatrix}0& e^{-\imath\varphi}\\
 e^{\imath\varphi}&0\end{bmatrix}
\end{align}
and $\widehat{V_{\alpha}^{K}}=(\widehat{V_{\alpha}^{K'}})^{*}$.

 In the one electron picture the
selection rules for optical transitions between LL $m\rightarrow
n$ are $\delta_{|m|,|n|\pm 1}$ \cite{Sadowski,Shon}.

\subsection{ Optical vectors for $\widetilde{H_{1<\nu<2}^{0}}$
and $\widetilde{H_{0<\nu<1}^{0}}$ }

In the case of $\widetilde{H_{1<\nu<2}^{0}}$ we obtain the
following components of $\overrightarrow{{\cal{M}}_{\parallel}}$
and $\overrightarrow{{\cal{M}}_{\perp}}$:

\begin{align}
\overrightarrow{{\cal{M}}_{\parallel}^{0}}&=
\frac{v_{F}}{\sqrt2}e^{-\frac{K^{2}}{4}}\{\sqrt{f_{0}^{+}},1,1,1,-\sqrt{1-f_{0}^{+}}\}
\nonumber\\
\overrightarrow{{\cal{M}}_{\perp}^{0}}&= \imath\frac{
v_{F}}{\sqrt2}e^{-\frac{K^{2}}{4}}\{\sqrt{f_{0}^{+}},1,1,-1,-\sqrt{1-f_{0}^{+}}\}
\end{align}
whereas for $0<\nu<1$, $f_{0}^{+}$ has to be replaced by
$f_{0}^{-}$.

It can be easily verified that for $K \simeq 0$ where all
transitions become degenerate, there is a sum rule such that
$\sum_{j}|
\overrightarrow{{\cal{M}}_{\parallel}}\cdot\overrightarrow{L_{j}}|^{2}=2
v_{F}^{2}$ independent of the filling factor whereas
$\sum_{j}(\overrightarrow{{\cal{M}}_{\parallel}}\cdot\overrightarrow{L_{j}})
(\overrightarrow{{\cal{M}}_{\perp}}\cdot\overrightarrow{L_{j}})^{*}=
\imath \nu v_{F}^{2}$. One therefore recover the selection rules
obtained for the one-electron model. Note that the Fermi velocity
entering in the optical matrix elements is that existing in the
\textit{absence} of electron-electron interactions.

\subsection{ Optical vectors for $\widetilde{HI_{1<\nu<2}^{12}}$ and
$\widetilde{HI_{0<\nu<1}^{12}}$ }

Following the same approach we get for the components of the
corresponding optical vectors
$\overrightarrow{{\cal{M}}_{\parallel}}$ and
$\overrightarrow{{\cal{M}}_{\perp}}$:

\begin{equation}
\begin{split}
\overrightarrow{{\cal{M}I}^{12}_{\parallel}}&=
\frac{v_{F}}{2}e^{-\frac{K^{2}}{4}}(1+
K^{2}\frac{\sqrt{2}-1}{2\sqrt{2}})\\
&\{1,-1,1,-1,1,-1,1,-1 \}\\
\overrightarrow{{\cal{M}I}^{12}_{\perp}}&=\imath
\frac{v_{F}}{2}e^{-\frac{K^{2}}{4}}(1-
K^{2}\frac{\sqrt{2}+1}{2\sqrt{2}})\\
& \{1,1,1,1,1,1,1,1 \}
\end{split}
\end{equation}

 which are no longer dependent of the filling factor for
 $\nu<2$. It can be shown that the only optical active transition
 are those corresponding to the solutions $EI_{1}^{+/-}(K)$ of Eq.
 18.

\subsection{ Optical vectors for $\widetilde{HI_{1<\nu<2}^{23}}$ and
$\widetilde{HI_{0<\nu<1}^{23}}$ }

In this case we get for the components of the corresponding
optical vectors $\overrightarrow{{\cal{M}}_{\parallel}}$ and
$\overrightarrow{{\cal{M}}_{\perp}}$:

\begin{equation}
\begin{split}
\overrightarrow{{\cal{M}I}^{23}_{\parallel}}&=
\frac{v_{F}}{2}e^{-\frac{K^{2}}{4}}(1+
(\frac{3}{\sqrt{6}}-1)K^{2}+(\frac{1}{4}-\frac{1}{\sqrt6})\frac{K^{4}}{2})\\
&\{1,-1,1,-1,1,-1,1,-1 \}\\
\overrightarrow{{\cal{M}I}^{12}_{\perp}}&=\imath
\frac{v_{F}}{2}e^{-\frac{K^{2}}{4}}(1-
(\frac{3}{\sqrt{6}}+1)K^{2}+(\frac{1}{4}+\frac{1}{\sqrt6})\frac{K^{4}}{2})\\
&\{1,1,1,1,1,1,1,1 \}
\end{split}
\end{equation}

\subsection{ Optical vectors for $\widetilde{H_{2<\nu<6}^{12}}$ }

Here one gets for the components of the corresponding optical
vectors $\overrightarrow{{\cal{M}}_{\parallel}}$ and
$\overrightarrow{{\cal{M}}_{\perp}}$ the following relations where
we have defined the functions
$p_{\parallel}(K)=\frac{1}{\sqrt2}(1-\frac{K^{2}(\sqrt2
+1)}{2\sqrt2})$ and
$p_{\perp}(K)=\frac{1}{\sqrt2}(1-\frac{K^{2}(\sqrt2
-1)}{2\sqrt2})$:

For $2<\nu<3$:
\begin{equation}
\begin{split}
\overrightarrow{{\cal{M}}_{\parallel}^{12}}_{2<\nu<3}&=
\frac{v_{F}e^{-\imath\varphi}}{\sqrt2}e^{-\frac{K^{2}}{4}}\\
&\{1,1,1,\sqrt{1-f_{0}^{-}},p_{\parallel}(K)\sqrt{f_{0}^{-}}\}\\
\overrightarrow{{\cal{M}}_{\perp}^{12}}_{2<\nu<3}&=
\imath\frac{v_{F}e^{-\imath\varphi}}{\sqrt2}e^{-\frac{K^{2}}{4}}\\
&\{1,1,1,\sqrt{1-f_{0}^{-}},p_{\perp}(K)\sqrt{f_{0}^{-}}\}
\end{split}
\end{equation}

For $3<\nu<4$:
\begin{equation}
\begin{split}
\overrightarrow{{\cal{M}}_{\parallel}^{12}}_{3<\nu<4}&=
\frac{v_{F}e^{-\imath\varphi}}{\sqrt2}e^{-\frac{K^{2}}{4}}\\
&\{1,1,\sqrt{1-f_{0}^{+}},p_{\parallel}(K),p_{\parallel}(K)\sqrt{f_{0}^{+}}\}\\
\overrightarrow{{\cal{M}}_{\perp}^{12}}_{3<\nu<4}&=
\imath\frac{v_{F}e^{-\imath\varphi}}{\sqrt2}e^{-\frac{K^{2}}{4}}\\
&\{1,1,\sqrt{1-f_{0}^{+}},p_{\perp}(K),p_{\perp}(K)\sqrt{f_{0}^{+}}\}
\end{split}
\end{equation}

For $4<\nu<5$:
\begin{equation}
\begin{split}
\overrightarrow{{\cal{M}}_{\parallel}^{12}}_{4<\nu<5}&=
\frac{v_{F}e^{-\imath\varphi}}{\sqrt2}e^{-\frac{K^{2}}{4}}\\
&\{1,\sqrt{1-f_{0}^{-}},p_{\parallel}(K)\sqrt{f_{0}^{-}},p_{\parallel}(K),p_{\parallel}(K)\}\\
\overrightarrow{{\cal{M}}_{\perp}^{12}}_{4<\nu<5}&=
\imath\frac{v_{F}e^{-\imath\varphi}}{\sqrt2}e^{-\frac{K^{2}}{4}}\\
&\{1,\sqrt{1-f_{0}^{-}},p_{\perp}(K)\sqrt{f_{0}^{-}},p_{\perp}(K),p_{\perp}(K)\}
\end{split}
\end{equation}

 For $5<\nu<6$:
\begin{equation}
\begin{split}
\overrightarrow{{\cal{M}}_{\parallel}^{12}}_{5<\nu<6}&=
\frac{v_{F}e^{-\imath\varphi}}{\sqrt2}e^{-\frac{K^{2}}{4}}\\
&\{\sqrt{1-f_{0}^{+}},p_{\parallel}(K)\sqrt{f_{0}^{+}},p_{\parallel}(K),p_{\parallel}(K),p_{\parallel}(K)\}\\
\overrightarrow{{\cal{M}}_{\perp}^{12}}_{5<\nu<6}&=
\imath\frac{v_{F}e^{-\imath\varphi}}{\sqrt2}e^{-\frac{K^{2}}{4}}\\
&\{\sqrt{1-f_{0}^{+}},p_{\perp}(K)\sqrt{f_{0}^{+}},p_{\perp}(K),p_{\perp}(K),p_{\perp}(K)\}
\end{split}
\end{equation}

It can be  verified that, for $K \simeq 0$, there is a sum rule
such that for the transitions $n=0$ to $n=1$, $\sum_{j}|
\overrightarrow{{\cal{M}}_{\parallel}}\cdot\overrightarrow{L_{j}}|^{2}=\frac{6-\nu}{2}
v_{F}^{2}$  whereas for the transitions $n=1$ to $n=2$, $\sum_{j}|
\overrightarrow{{\cal{M}}_{\parallel}}\cdot\overrightarrow{L_{j}}|^{2}=\frac{\nu-2}{4}
v_{F}^{2}$.

We also get
$\sum_{j}(\overrightarrow{{\cal{M}}_{\parallel}}\cdot\overrightarrow{L_{j}})
(\overrightarrow{{\cal{M}}_{\perp}}\cdot\overrightarrow{L_{j}})^{*}=
\imath \sum_{j}|
\overrightarrow{{\cal{M}}_{\parallel}}\cdot\overrightarrow{L_{j}}|^{2}$.
One therefore recovers the selection rules obtained for the
one-electron model.

To conclude this part, it is worth comparing these results with
those obtained in  C2DEG where the introduction of
electron-electron interactions change the selection rules
\cite{By2,Faugeras} with respect to those obtained in the
one-electron picture. In the case of Graphene, the selection rules
are in general, similar to the results obtained in the
one-electron picture, except that the oscillator strength is
condensed into one or two branches of the MP curves and that the
strength remains proportional to $v_{F}^{2}$ and not to the square
of the re-normalized velocity.
\\


\end{document}